\documentstyle[12pt,moriond]{article}
\begin{document}

\heading{Properties of High Redshift Galaxies }

\author{D. Calzetti} {Space Telescope Science Institute, 
3700 San Martin Drive, Baltimore, MD 21218, U.S.A.} 

\begin{moriondabstract}
I review the characteristics of high redshift galaxies, with
particular attention to the effects of dust obscuration on the
observed light. Galaxies at redshift z$\sim$1 and at z$>$2 are
discussed separately, as the accessible information for each redshift
range are different. In the z$\le$1--2 redshift region, data on the
H$\alpha$ and mid/far-IR luminosity of the galaxies are becoming
increasingly available, effectively complementing the restframe UV
data; they offer a better handle on the amount of dust obscuration
present in the galaxies, although some ambiguity remains as to the
distribution of the dust within the system and the opacity correction
factor this implies. At higher redshifts, most of the available data
cover the far-UV region only, representing a challenge for a careful
assessment of the effects of dust in galaxies. Finally, I discuss the
implications of the dust corrections to the star formation rate
density of the Universe as a function of redshift.
\end{moriondabstract}

\def\putplot#1#2#3#4#5#6#7{\begin{centering} \leavevmode
\vbox to#2{\rule{0pt}{#2}}
\includegraphics{#1}
\end{centering}}
%

\section{Introduction}

The study of galaxy populations at high redshift is aimed at
understanding how galaxies evolved into their low-redshift
counterparts. One of the tools for achieving this goal is the
investigation of the stellar populations, their evolution, and their
relation to the galaxy morphological type, via the observation and
analysis of the emitted light.

Current instrumentation provides most immediate access to the optical
and near-IR part of the electromagnetic spectrum, which correspond to
restframe UV and optical wavelengths for redshifts z$\ge$1. The
availability of restframe UV measurements is particularly important
for quantifying the evolution of the star formation rate in galaxies
(Lilly et al.  1996, Madau et al 1996). Newly formed massive stars
emit the bulk of their energy in the UV (e.g., Leitherer \& Heckman
1995), and light in this waveband is a sensitive indicator of 
recent star formation, over timescales of $\sim$100~Myr. Measuring the
evolution of the star formation is a direct way to measure the
evolution of the stellar populations in a galaxy.

The UV emission from a galaxy is, however, heavily affected by the
presence of even small quantities of dust. Sources which are optically
thin in the visible (e.g., A$_V$=0.3) can be optically thick in the UV
(A$_{1300 \AA}\simeq$1). Light at longer wavelengths is progressively
less sensitive to the effects of dust obscuration, but also less
sensitive to recent star formation (if we exclude the Red Supergiants
in the near-IR). Exception to this rule are the nebular emission lines
(e.g., H$\beta$, H$\alpha$, [OII], etc.), which are directly excited
by ionizing photons, and the emission from dust at wavelengths
$\ge$30~$\mu$m. Redshifted nebular emission lines fall in the
near/mid-IR wavelength range; for instance, the bluest of the strong
emission lines, [OII]($\lambda$3727~\AA) is redshifted to 1.1~$\mu$m
in a z=2 galaxy. The detection and measurement of emission lines in
high redshift galaxies is rather difficult at present because of the
limited sensitivity of near-IR and mid-IR instrumentation (see,
however, the results of Pettini et al. 1998, Glazebrook et al. 1999,
Yan et al. 1999). Dust heated by hot stars to temperatures T$\ge$30~K
emits in the far-IR and represents an indirect, but sensitive,
indicator of star formation in a galaxy (e.g., Helou 1986, Young et
al. 1989); in addition, it is, by definition, insensitive to dust
obscuration. Recent far-IR and sub-mm instruments, like ISO and SCUBA,
have reached high enough sensitivities to be able to measure dust
emission from distant galaxies. There are, nevertheless, residual
problems with this type of measures: limitations by source confusion,
uncertainties in source positioning which hamper the assignment of
optical counterparts and, therefore, of redshifts to the dust-emitting
sources, possibility of AGN contamination, and uncertainties in the
dust spectral energy distribution (SED) represent difficulties which
have not been completely overcome yet (see, however, Barger et
al. 1999). The latter of these problems is briefly discussed in
section~3 below.

In this talk I concentrate on the UV and optical properties of
galaxies at redshift z$\ge$1, and the effects of dust obscuration
on the emerging light. Galaxies at z$\le$1 are discussed in another
contribution to these Proceedings (see, F. Hammer), while the
characteristics of medium--high redshift galaxy populations detected
in the far-IR and sub-mm are presented by S. Lilly, D. Hughes,
J.-L. Puget and others (also these Proceedings).

\section{Galaxies at Redshift z$\sim$1}

The restframe UV luminosity density of galaxies displays a
five/ten--fold increase from z$\sim$0 to z$\sim$1 (Lilly et
al. 1996). This result has been interpreted as a comparable increase
in the star formation rate density (SFRD) of the Universe with
redshift (Madau et al. 1996, and subsequent papers). Galaxies were
therefore forming stars at a higher pace about 9~Gyr ago than today,
roughly the same trend expected in CDM galaxy evolution models (e.g.,
Baugh et al. 1998). Although the observed trend is probably correct,
the actual numbers for the absolute value and the slope of the SFRD--z
relation derived from the restframe UV data needed independent
confirmation, because UV measurements are potentially affected by dust
obscuration. Nevertheless, it is likely that the intrinsic SFRD between
z=0 and z=1 is at most a factor $\sim$3--3.5 higher than what derived
from the UV measurements; larger values of the SFRDs would make
galaxies quickly run out of gas, under most assumptions for the
stellar IMF and for the gas recycling from supernovae and massive
stars (Calzetti \& Heckman 1999; Pei, Fall \& Hauser 1999). Recent ISO
15~$\mu$m observations of the galaxies in the Lilly et al. fields
confirm the large increase of the SFRD with redshift in the range
0.3--1, but also show that the absolute value of the SFRD can be as
much as $\sim$3 times higher than that measured from the UV data
(Flores et al. 1999, see also Hammer, this Conference). Still, we
should bear in mind that the derivation of the SFRD from the ISO data
is somewhat uncertain, as the extrapolation from the restframe
$\sim$10~$\mu$m flux density to the bolometric far-IR dust emission is
a non-trivial step.

\begin{figure}[h]
\putplot{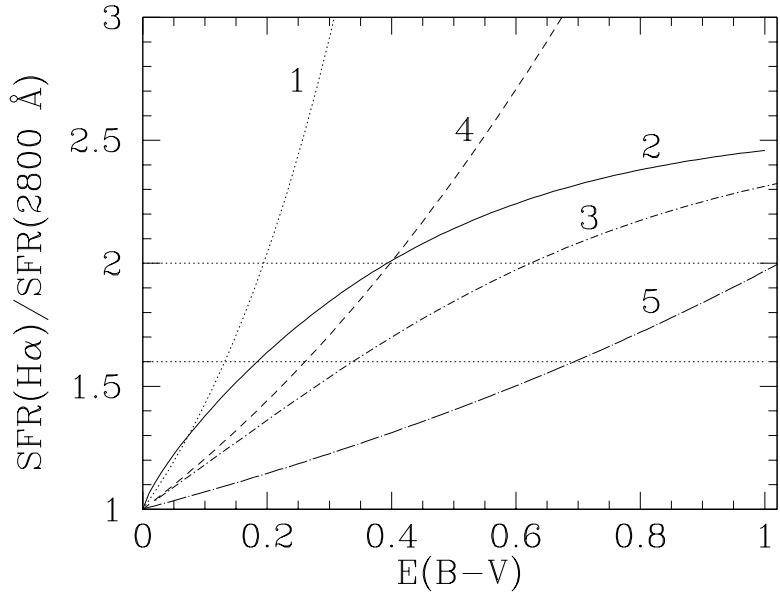}{7.0 cm}{-90}{40}{40}{100}{230}
\caption{The ratio of the star formation rates measured from the {\bf
observed} H$\alpha$ line emission and from the UV continuum emission
at 2800~\AA, SFR(H$\alpha$)/SFR(2800), plotted as a function of the 
mean galaxy color excess E(B$-$V), for 5 different dust geometry models: 
1. homogeneous, foreground screen; 2. homogeneous disk; 3. homogeneous 
spheroid; 4. clumpy shell; 5. starburst dust distribution (see text for 
explanations on the different models). The horizontal lines mark 
approximately the range given by the observational data (equation~1).}
\end{figure}

Galaxies at z$\le$1--1.5 can be tested for the presence of dust
obscuration by complementing the UV data with measurements at redder
wavelengths of, e.g., nebular emission lines, like H$\alpha$, which
are excited by young massive ionizing stars, but are also less
sensitive to the effects of dust.  Technically, this method only tests
for {\em dust reddening}, not dust obscuration, because regions which
are opaque at $\sim$7000~\AA~ will not be detected in either UV or
H$\alpha$.  Glazebrook et al. (1999) and Yan et al. (1999) measured
the H$\alpha$ emission from galaxies at redshift z$\approx$1. In both
cases it is found that the average SFR derived from H$\alpha$,
SFR(H$\alpha$), is systematically higher than the average SFR derived
from the restframe UV emission at 2800~\AA, SFR(2800). In particular:
\begin{equation}
{SFR(H\alpha)\over SFR(2800)} \sim 1.6 - 2,
\end{equation}
(with fairly large error bars). In the absence of dust obscuration,
SFR(H$\alpha$)/SFR(2800)=1, where reference SFRs are derived from
models of stellar populations with constant star formation since 1~Gyr
and a standard Salpeter stellar IMF (Glazebrook et al. 1999).  A value
$>$1 in equation~1 is expected if the emerging light is attenuated by
dust. It is to be remarked that the H$\alpha$ line emission and the UV
continuum emission at 2800~\AA~ are sensitive to {\em different} star
formation timescales: H$\alpha$ probes only the most recent star
formation, being produced only when the short-lived ionizing stars are
present (timescales of $\sim$10~Myr); the UV emission at 2800~\AA~ is
contributed also by long-lived, non-ionizing stars, and measures star
formation over timescales of $\sim$100--500~Myr. Therefore, if the
adopted model for the stellar population does not match reality, the
intrinsic value of the ratio SFR(H$\alpha$)/SFR(2800) can change by as
much as $\sim$30--60\%; for a population younger than $\sim$1~Gyr, the
change goes in the direction of decreasing the ratio of equation~1,
implying lower dust attenuations. For sake of simplicity, I assume in
the following that a 1~Gyr constant star formation stellar population
is a fair representation of a z$\sim$1 galaxy.

\begin{figure}[h]
\putplot{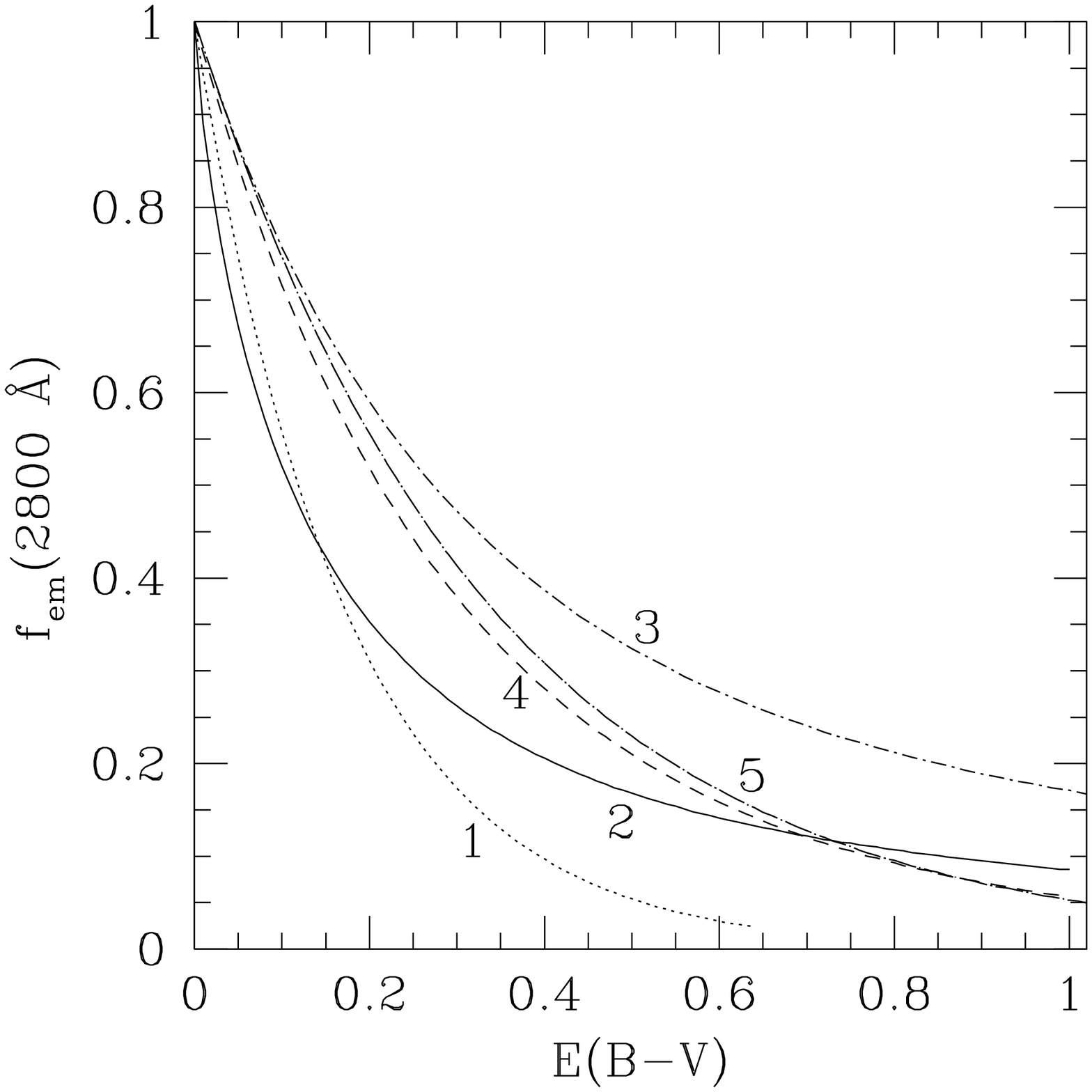}{3.2 cm}{0}{41}{41}{-10}{-180}
\putplot{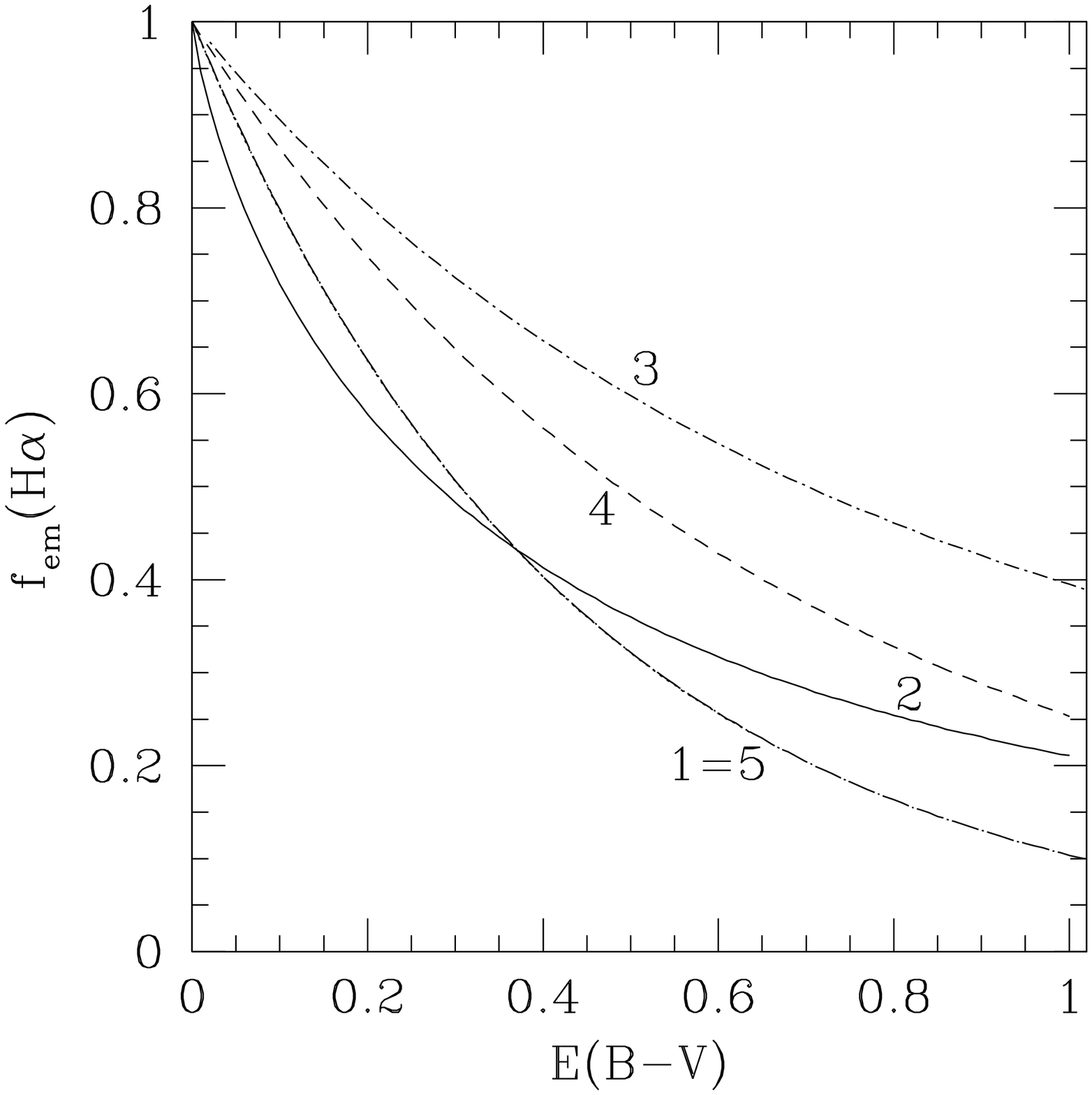}{3.2 cm}{0}{40}{40}{240}{-68} 
\caption{The fraction of emerging light in the UV continuum (left panel) 
and in the H$\alpha$ line (right panel) as a function of increasing 
mean galaxy color excess E(B$-$V). The numbering of the five dust models 
used to derive f$_{em}$ follows the same convention as Figure~1.}
\end{figure}

Establishing that the z$\sim$1 galaxies are affected by dust reddening
is far more straightforward than measuring {\em the amount of dust
reddening} itself. This measurement involves making assumptions on the
distribution of the dust inside the galaxy, and quite different
results can be obtained for different dust geometries (Witt, Thronson
\& Capuano 1992, Calzetti, Kinney \& Storchi-Bergmann 1994). Examples
are shown in Figures~1 and 2. In these figures, five simple models for
the geometry of dust in galaxies are used to derive the observed
SFR(H$\alpha$)/SFR(2800) and the fraction of light emerging in the UV
continuum and in the H$\alpha$ line as a function of the dust optical
depth, here given as color excess E(B$-$V). The first model is the
standard homogeneous, foreground dust screen, frequently used to
derive dust reddening in galaxies (e.g., Glazebrook et al. 1999, Yan
et al. 1999); this model is an extreme case under most assumptions,
since it implies not only that all the dust is completely foreground
to all the stars in the galaxy, but also that the dust screen is far
away enough from the galaxy that scattering of light into the line of
sight is a negligible effect (e.g., Calzetti et al. 1994). In the
second model the dust and the stars are homogeneously mixed together
and are distributed in a flattened disk (homogeneous disk); the case
shown in Figures~1 and 2 assumes that the opacity is averaged over all
possible inclination angles of the disks. The third model is also an
homogeneous mixture of dust and stars, but the two are distributed in
a sphere (homogeneous spheroid). In the fourth model, the stars are
centrally concentrated, while the dust is clumpy and located in a
shell surrounding the stars (clumpy shell). The fifth model uses the
distribution of dust and stars typically found in starburst galaxies
at low redshift (Calzetti et al. 1994, Calzetti 1997a, Calzetti et
al. 1999); here the distribution of the dust is roughly equivalent to
a clumpy shell (Gordon, Calzetti \& Witt 1997), except that the
ionized gas is on average twice as attenuated as the stellar continuum
(Calzetti et al. 1994).

Figure~1 shows that the same value of SFR(H$\alpha$)/SFR(2800)
corresponds to rather different values of the color excess in
different dust models.  For instance, for the homogeneous disk 
(model~2), SFR(H$\alpha$)/SFR(2800)=2 implies E(B$-$V)=0.4, with 21\%
and 41\% of the UV and H$\alpha$ light, respectively, emerging from
the galaxy. For the starburst dust distribution (model~5), the same
SFR ratio implies E(B$-$V)=1.0 and 5\% and 10\% of the UV and
H$\alpha$ light, respectively, emerging from the galaxy.

If we exclude the homogeneous screen model, the observed range of
SFR(H$\alpha$)/SFR(2800) implies correction factors between 2.3 and 5
for the UV light at 2800~\AA~ and between 1.4 and 2.4 for the
H$\alpha$ line, for models~2 through 4. A more extreme situation is
represented by the starburst reddening model, where only between 5\%~
and 13\% of the UV light and between 10\% and 21\% of the H$\alpha$
light emerge from the galaxy. This extreme behavior is due to the
differential reddening between ionized gas and stellar continuum. A
dust correction factor 10 or so for the observed UV flux density would
give, however, an excessively high value of the intrinsic SFRD at
z$\approx$1, probably high enough to deplete galaxies of their gas
content before z=0. Thus, the starburst model does not appear
applicable to the z$\sim$1 galaxies; this is expected for systems
older than a few Gyrs, which have evolved beyond their initial few
generations of stars, and whose emission is not dominated by a central
starburst.

The basic conclusion is that, if the values of equation~1 are
confirmed by further observations, the intrinsic SFRD at z$\sim$1 is
more than a factor 2 and likely less than a factor 5 larger than what
directly inferred from the observed UV flux density.

\section{Galaxies at Redshift z$>$2}

The population of galaxies at z$>$2--2.5 identified with the
Lyman-break technique (Steidel et al. 1996) show observational
characteristics similar to those of the central regions of local
UV-bright starburst galaxies. In both cases the objects are active
star--forming systems with observed SFRs per unit area of
$\approx$1~M$_{\odot}$~yr$^{-1}$~kpc$^{-2}$ (Calzetti \& Heckman 1999,
Meurer et al. 1997).  Restframe UV spectra, which generally cover the
range 900--1800~\AA, show a wealth of absorption features, and
sometimes P-Cygni profiles in the CIV(1550~\AA) line (cf. the figures
in Steidel et al. 1996), typical of the predominance of young, massive
stars. Restframe optical spectra available for a few Lyman-break
galaxies (Pettini et al. 1998) show nebular emission lines with the
typical intensities seen in local starbursts (Meurer, Heckman,
\& Calzetti 1999).

Another characteristic the two populations share is the large spread
in restframe UV colors (e.g., Dickinson 1998). If the observed UV
stellar continuum is parametrized by a power law,
F($\lambda$)$\propto\lambda^{\beta}$, in the range
$\sim$1200--1800~\AA, Lyman-break galaxies cover a large range of
$\beta$ values, roughly from $-$3 to 0.4, namely from very blue to
moderately red (Dickinson 1999, private communication). This range is
not very different from that covered by the local, UV-bright
starbursts. Population synthesis models (e.g. Leitherer \& Heckman
1995) indicate that a dust-free, young starburst or constant
star-formation population have invariably values of $\beta<-$2.0, for
a vast range of metallicities. For the local starbursts, the most
straightforward intepretation for the observed range of UV spectral
indices is dust reddening (Calzetti et al. 1994). Dust reddening has
been proposed as an explanation also in the case of the Lyman-break
galaxies (Calzetti 1997b, Meurer et al. 1997, 1999, Pettini et
al. 1998, Steidel et al. 1999). The observed UV emission from
the galaxies at z$\sim$3 then accounts, on average, for
$\sim$20--25\%~ of the intrinsic UV luminosity (Steidel et
al. 1999). Similar fractions of observed-to-intrinsic UV luminosity
are expected in evolution models of the dust content of galaxies
(Calzetti \& Heckman 1999, Pei et al.  1999).

\begin{figure}[h]
\putplot{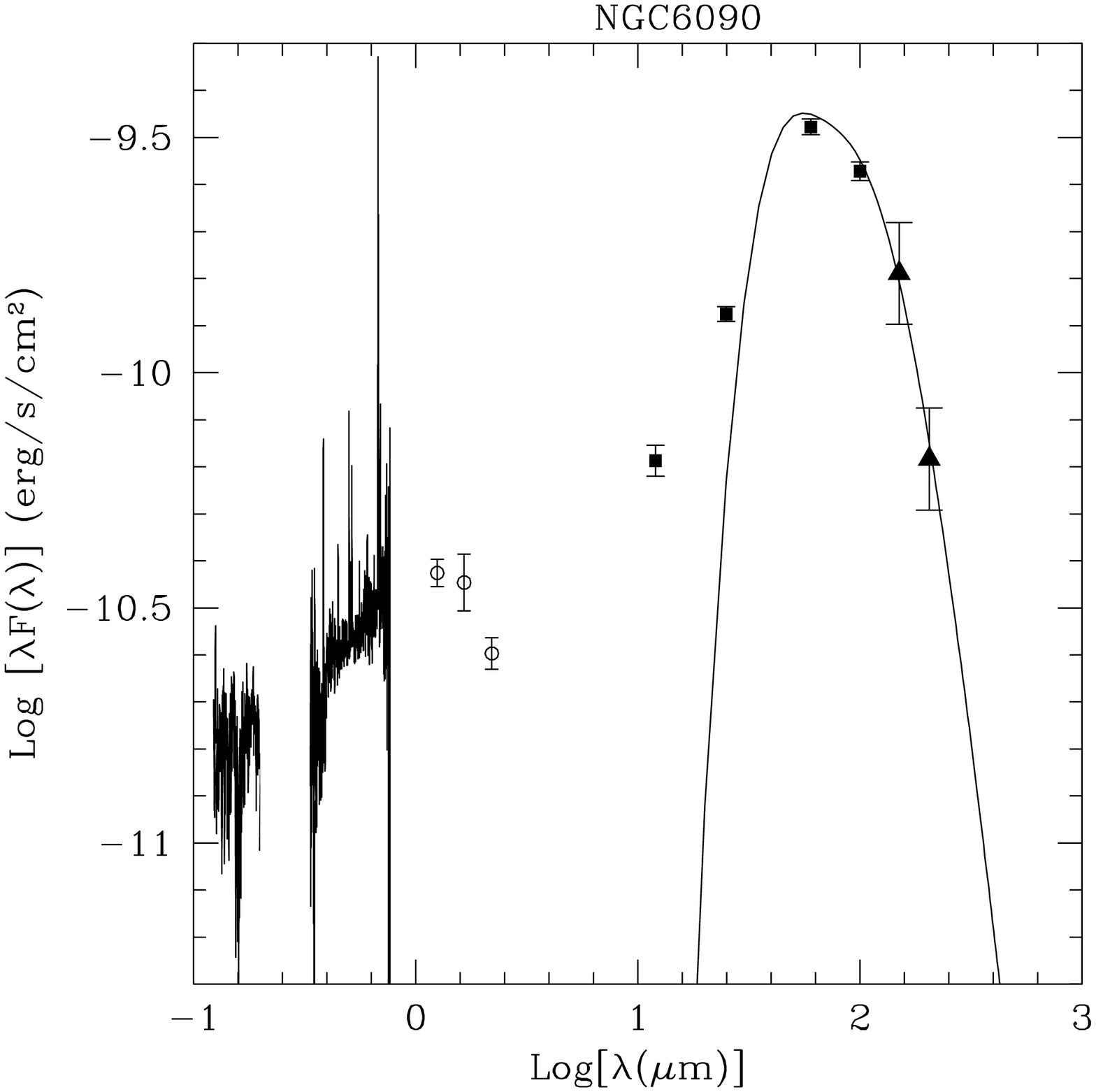}{3.2 cm}{0}{41}{41}{-10}{-180}
\putplot{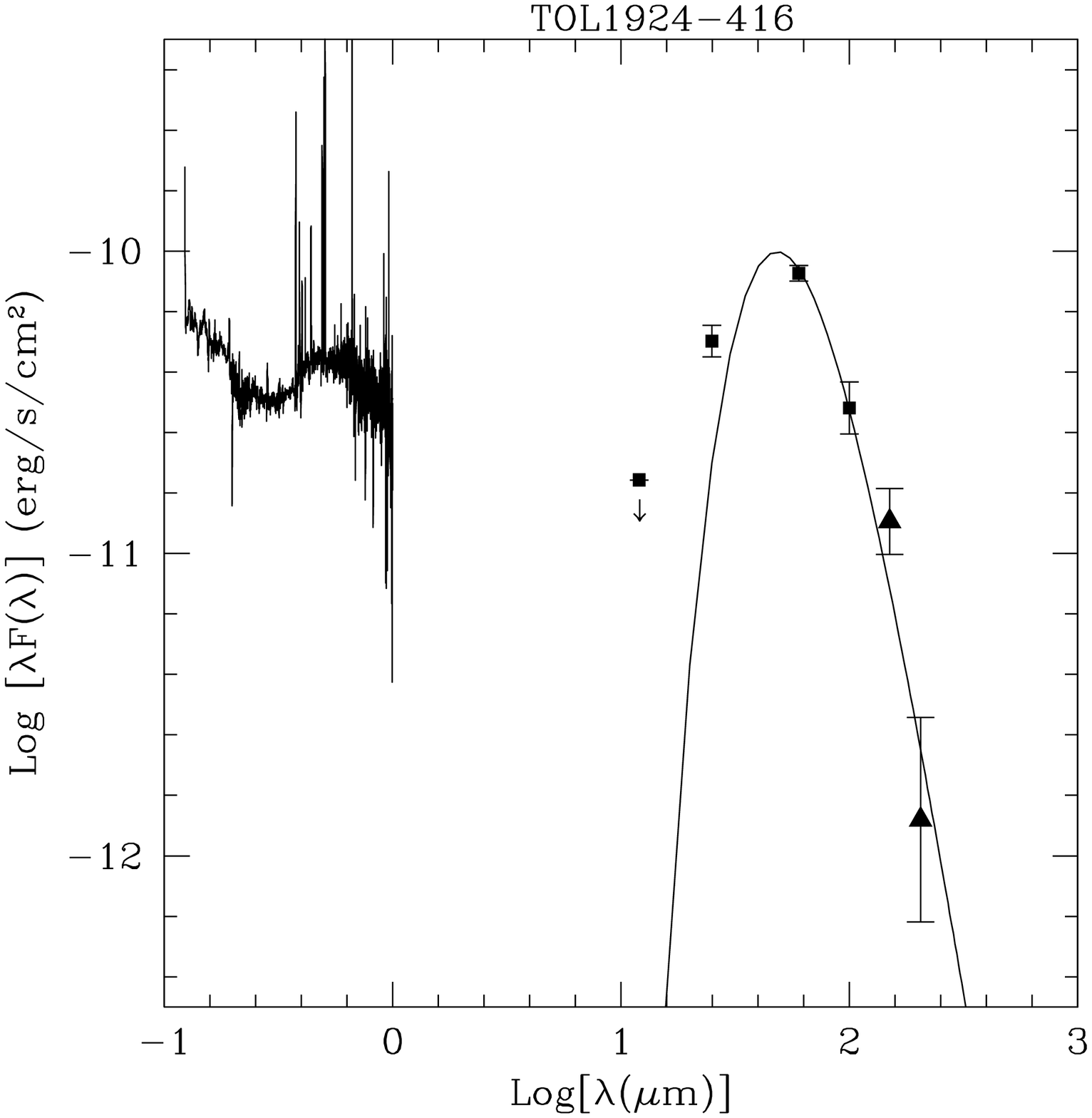}{3.2 cm}{0}{40}{40}{240}{-68} 
\caption{The UV-optical-nearIR-farIR SEDs of two low-redshift
starburst galaxies: the FIR-luminous NGC6090 (left panel) and the Blue
Compact Dwarf Tol1924$-$416 (right panel). UV--optical
spectrophotometry is shown as a continuous line in the wavelength
range 0.12-1.0~$\mu$m, nearIR photometry as empty circles, IRAS data
as filled squares, and ISO measurements as filled triangles (Calzetti
et al. 1999). Upper limits have downward arrows. For NGC6090, the
far-IR data are fit by two modified Planck functions with temperatures
T$_1$=49~K and T$_2$=23~K and, for Tol1924$-$416, by a single modified
Planck function with temperature T=50~K. The fits are shown as solid
curves.  The emissivity of the dust is taken to be $\nu^2$. }
\end{figure}

If the interpretation of the high-z UV color spread is correct, the
light absorbed by dust at UV--optical wavelengths should appear in the
far-IR as dust emission. As already discussed in the previous section
for UV and H$\alpha$, the UV slope $\beta$ technically measures dust
reddening, not dust obscuration; i.e., UV light from stars deeply
buried inside dust does not contribute, in principle, to the UV
emission and slope (Calzetti et al. 1995).  Meurer et al. (1999)
tested the relation between reddening and obscuration in local
starburst galaxies by comparing $\beta$ with FIR$_{IRAS}$/F(UV); this
ratio is a measure of total dust obscuration because FIR$_{IRAS}$ and
F(UV) have roughly opposite trends for increasing amounts of dust
(FIR$_{IRAS}$ increases while F(UV) tends to decrease), while they are
both proportional to the SFR of the galaxy. Calzetti et al. (1999) use
recent ISO observations of local starbursts to directly compare the
amount of energy emitted by dust in the far-IR with predictions for the
energy absorbed by dust in the UV--optical; the latter quantity is
derived using the reddening recipe of Calzetti et al. (1994) and
Calzetti (1997b). The basic conclusion is that reddening and total
obscuration are correlated in local, UV-bright starburst galaxies; the
recipe of Calzetti (1997b) can then be used to predict the total
amount of energy absorbed by dust, provided that the value
R$^{\prime}_V$=4.05 (instead of 4.88, see Calzetti 1997b) is used in
conjunction with the reddening curve. The predicted values match the
observed far-IR emission of individual cases within a factor 2, but,
when averaged over a large sample of galaxies, the predicted value
matches the median of the observations within 20\%.

As said above, the Lyman-break galaxies resemble local UV-bright
starbursts; the reddening/obscuration recipe derived for the latter
objects should then be applicable to the former ones to obtain
predictions on the total far-IR dust emission (Meurer et al. 1999, Calzetti
et al. 1999). However, whether a high-redshift galaxy will be
detectable or not with the currently available instrumentation depends
not only on the total far-IR flux but also on the dust SED.  SCUBA at
the JCMT has been proven very effective at detecting galaxies at
cosmological distances, especially in the sensitive 850~$\mu$m band
(e.g., Hughes et al. 1998, Eales et al. 1998, Blain et al. 1999, Lilly
et al. 1999, Barger et al. 1999). For a z$\sim$3 galaxy, this band
probes the restframe $\sim$210~$\mu$m emission, longward of the
peak of the dust SED, usually located between 60 and
100~$\mu$m in actively star-forming galaxies (e.g., Helou 1986, see
Figure~3). The fraction of light emitted at 200~$\mu$m relative to the
total far-IR dust luminosity varies by at least one order of magnitude
among starbust galaxies. An example is shown in Figure~3, where the
UV--to--far-IR SEDs of two local starburst galaxies are reported. In the
FIR-luminous galaxy NGC6090, 
$\lambda$F($\lambda$)$_{205}$/F$_{far-IR}$=0.12, where
$\lambda$F($\lambda$)$_{205}$ is the energy coming out in the
205~$\mu$m ISO band and F$_{far-IR}$ is the total far-IR dust
emission; in the Blue Compact Dwarf galaxy Tol1924$-$416, 
$\lambda$F($\lambda$)$_{205}$/F$_{far-IR}$=0.013, about a factor 10
smaller than in NGC6090 (Calzetti et al. 1999). Indeed, Tol1924$-$416 has
hotter dust than NGC6090. If the far-IR SED of each galaxy is fitted
with two modified Planck functions at different temperatures, the two
temperatures have values T$_1$=49~K and T$_2$=23~K for NGC6090, but
only one temperature value, T=50~K, can be derived for
Tol1924$-$416. This galaxy is missing the cooler dust component (a
result independent of the adopted dust emissivity index).

Why is Tol1924$-$416 warmer than NGC6090? One possible reason is 
the difference in the total dust content of the two galaxies: the Blue Compact
Dwarf is about 10 times more metal poor than the FIR-luminous NGC6090,
implying that, for the same gas content, the dust column density is
also about 10 times smaller.  Dust self-shielding is less efficient in
the metal-poor object and the dust will tend to be hotter (Mezger,
Mathis \& Panagia 1982).

Detecting dust emission in just one band thus leaves the ambiguity of
the actual SED to be adopted to derive the total far-IR emission from
the galaxy. Observations at sub-mm and mm wavelengths in multiple
bands will be needed to actually understand the physical
characteristics of the far-IR emission from high-redshift galaxies
(e.g., is the dust in UV-bright Lyman-break galaxies hotter than in local
galaxies?), and pin down the total amount of dust contained in the 
distant progenitors of present-day galaxies.

\section{The Evolution of the Star Formation Rate Density}

In recent years, a wealth of data have become available on the
luminosity density of the Universe at a variety of wavelengths and
redshifts. One of the scopes of these measurements is to derive the
evolution of the SFRD of the Universe (see the pioneering work of
Madau et al. 1996), as a way to characterize the global evolution of
the stellar populations in galaxies.

A number of uncertainties affect the conversion of a luminosity
density into a SFRD, in addition to the problems usually encountered
in statistical studies, such as volume corrections, luminosity
selections, etc. A throrough discussion of all the uncertainties is
given by D. Schaerer (these Proceedings); a few of those relevant to
the present discussion are mentioned here. For UV measurememts, the
largest uncertainty is represented by dust corrections; as seen in the
previous sections, presence of dust in galaxies can absorb between
50\% and 80\% of the UV light. The next largest uncertainty at all
wavelengths is the stellar IMF, which is poorly known especially at
the low mass end, where most of the stellar mass is contained; this
carries a factor 2--3 uncertainty in the conversion of a luminosity
density into a SFRD. For measurements longward of $\sim$2,000~\AA~
(excluding recombination lines), another source of uncertainty is the
galaxy star formation history: the longer a galaxy has been forming
stars, the more intermediate/low mass stars have been accumulating,
and the more flux is accumulating in the integrated light at redder
wavelengths. Variations of the star formation history between galaxies
induces uncertainties up to $\sim$30--50\%~ in the SFRD derived from,
e.g., data at 2800~\AA.

Figure~4 reports UV, H$\alpha$, and mid/far-IR luminosity density data
at a variety of redshifts, collected from the literature; the
luminosities have been converted to SFRDs assuming a Salpeter IMF with
mass range 0.35--100~M$_{\odot}$ and a 1~Gyr constant star formation
stellar population. The data form a `progression' from UV to H$\alpha$
to mid/far-IR in the left panel of Figure~4: the longer the
wavelength, the higher, on average, the estimated SFRD. This is
consistent with presence of dust obscuration in the galaxies. The
right panel of Figure~4 reports the same data after dust correction:
the spread at each redshift value is now smaller than in the left
panel.  The UV and H$\alpha$ data have been corrected according to the
results reported in the previous sections, while the mid/far-IR data
have not been corrected. For z$<$1, I have assumed that the
corrections to apply to the UV data are the same as those derived for
the z$\sim$1 galaxies. In addition, among the H$\alpha$ luminosity
points, only the data of Glazebrook et al. (1999) and Yan et
al. (1999) have been corrected (the point from Gronwall 1998 has been
already corrected by the author).

\begin{figure}[h]
\putplot{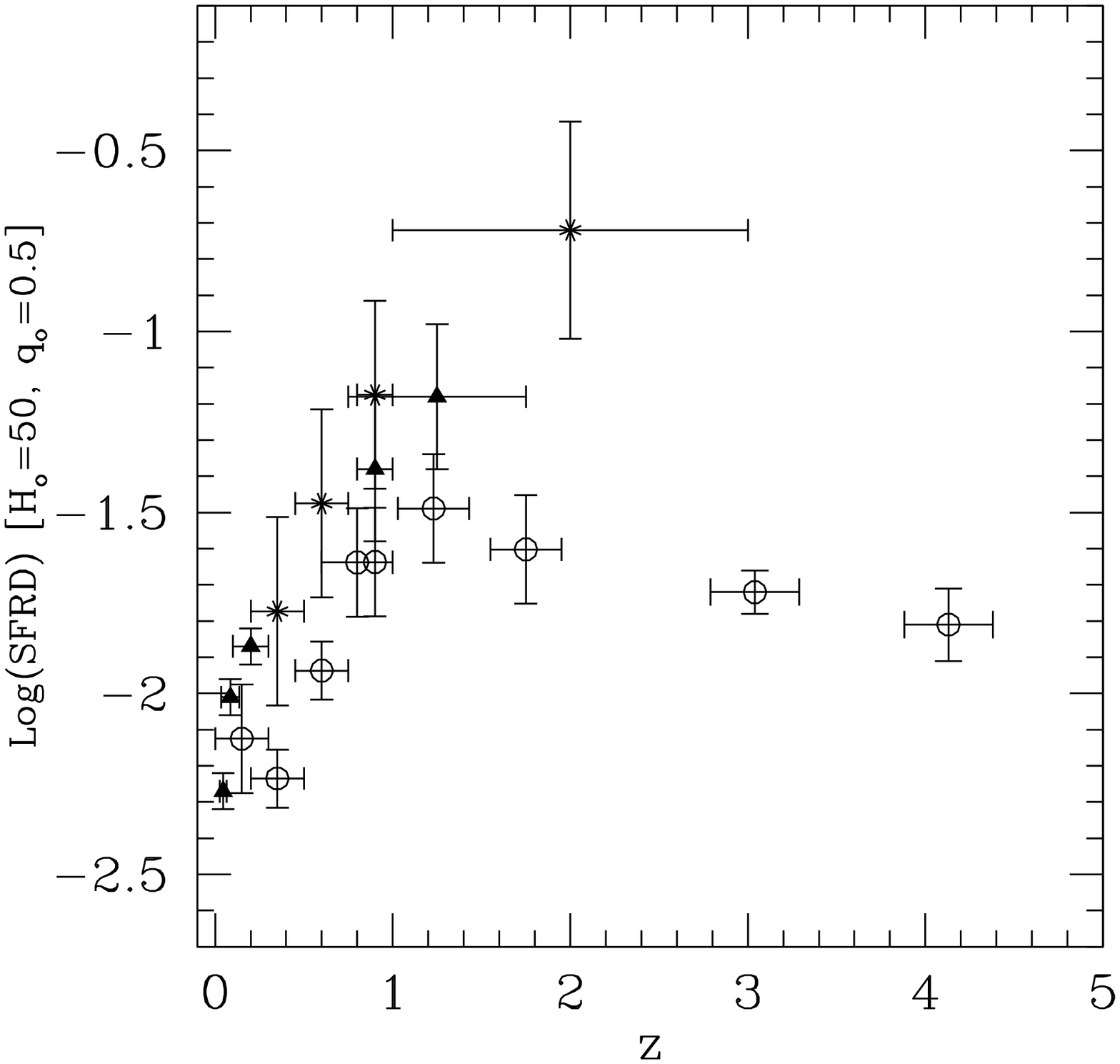}{3.2 cm}{0}{41}{41}{-10}{-180}
\putplot{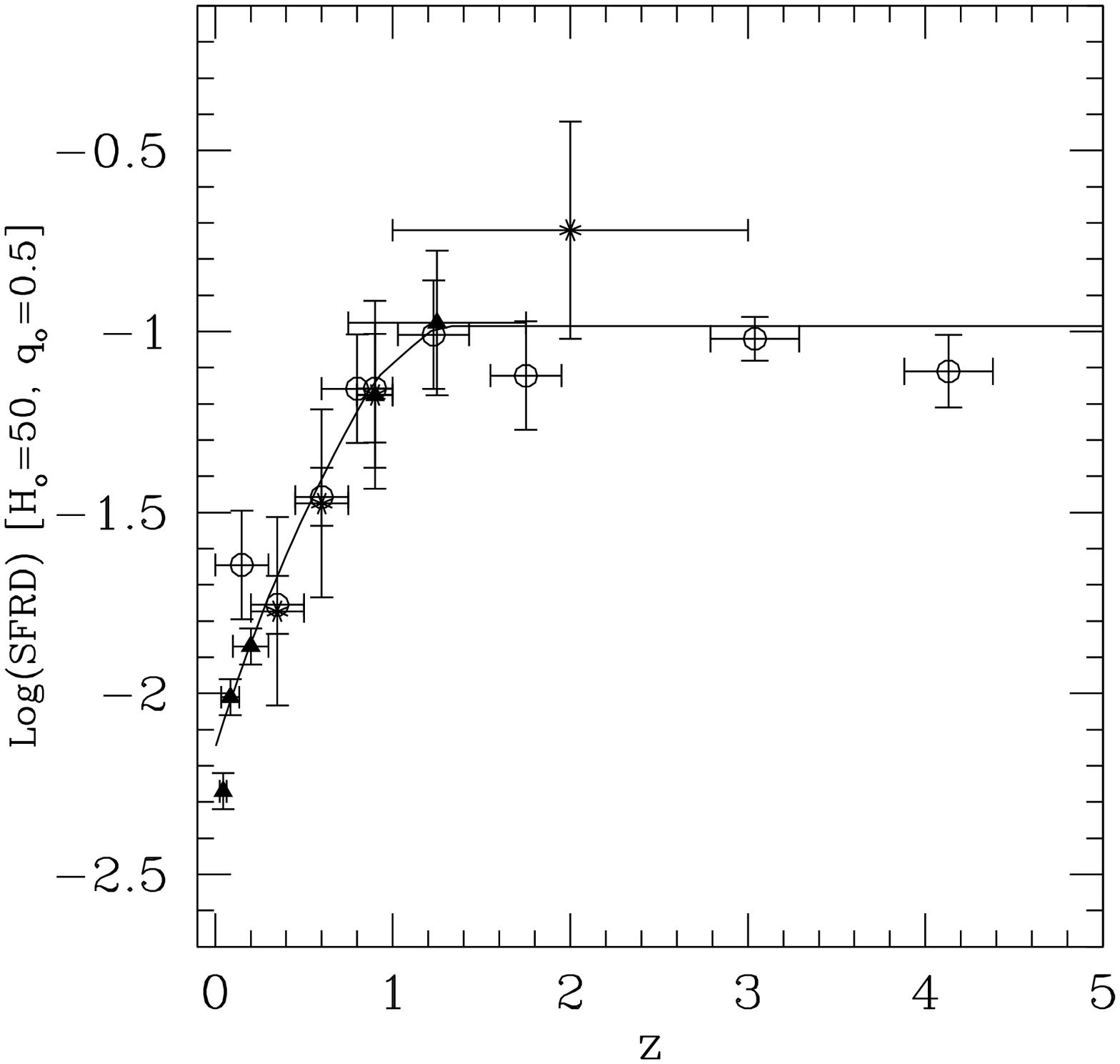}{3.2 cm}{0}{40}{40}{240}{-68} 
\caption{The average star formation rate density in the Universe,
SFRD, in M$_{\odot}$~yr$^{-1}$~Mpc$^{-3}$ as a function of redshift,
before (left panel) and after (rigth panel) correction for the effects
of dust obscuration. Empty circles indicate restframe UV data at
2800~\AA~ (for z$<$2, Lilly et al. 1996, Connolly et al. 1997) and at
1700~\AA~ (for z$>$2, Steidel et al. 1999). Filled triangles mark the
position of the H$\alpha$ measurements (Gallego et al. 1995, Gronwall
1998, Tresse \& Maddox 1998, Glazebrook et al. 1999, Yan et
al. 1999). Asterisks mark the position of mid/far-IR data, from ISO
(z$<$1, Flores et al. 1998) and from SCUBA (see text for
references). 1~$\sigma$ error bars are reported for all the data. The
continuous line in the right panel represents one of the models for
the evolution of the SFRD of the Universe from Calzetti \& Heckman
(1999, their Model~3).}
\end{figure}

The dust-corrected data of Figure~4 are consistent with the SFRD being
roughly constant between z$\sim$1.2--1.4 and z$\sim$4 and about 5
times larger than what directly measured from UV data. Between
z$\sim$1.2 and z=0, the SFRD decreases by a factor $\sim$10 (Lilly et
al. 1996) and is about 3 times larger than what inferred from the UV
measurements. The comparison with one of the SFRD evolution models of
Calzetti \& Heckman (1999) highlights the trend just described. This
implies that up to $\sim$20\% of all the stars were formed before z=3
and up to $\sim$60\% before z$\sim$1.2. Excess of metal production has
often been put forward as an argument against large SFRD values at
high redshifts; however, gas inflows/outflows in/out of galaxies can
completely offset the problem, as the excess metals get dispersed into
the IGM. One of the advantages of a high SFRD at z$>$2 is the ability
to reproduce the Cosmic Infrared Background observed by COBE (see, for
the observations, Fixsen et al. 1998, Hauser et al. 1998, and, for the
models, Calzetti \& Heckman 1999, Pei et al. 1999).

In conclusion, dust obscuration affects the observed light from high
redshift galaxies, with flux reductions up to a factor 5 when
targeting the restframe UV. This happens even and especially at high
redshift. Although common wisdom would suggest that metallicities are
low in the young Universe, gas column densities are high so dust
opacity can be high; in addition the observer optical waveband
corresponds to restframe far-UV, a wavelength region strongly affected
by dust obscuration. 

One question which has not been answered yet, but is very important
for measuring the SFRD at high redshift, is the relationship between
the UV-bright Lyman-break galaxies and the FIR-bright SCUBA galaxies.
In Figure~4 the data from the Lyman-break galaxies and from the SCUBA
galaxies are used as if originating from the same objects (although at
different wavelengths). However, if the two type of galaxies are
complementary, rather than overlapping, populations, their
contributions to the SFRD would sum up, rather than average together,
increasing the high redshift SFRD by another factor $\sim$2. In this
case, a problem may arise, as the global SFR would be large enough to
deplete galaxies of their gas content before z=0. Clearly, the nature
of the two galaxy populations need to be addressed, to understand
whether such a problem may exist.

\section{Acknowledgements}

I would like to thank C. Leitherer (STScI) for a critical reading of the 
manuscript and the Organizing Committe of the XXXIVth Recontres
de Moriond for inviting me to this very interesting meeting and for
financially supporting my stay at Les Arcs. Part of this trip has been
supported by the STScI Director Discretionary Research Funds.


%
\vfill
\end{document}